# A Practical Guide for Supporting Formative Assessment and Feedback Using Generative AI

Sapolnach Prompiengchai[1,2,3*], Charith Narreddy[3,4], Steve Joordens[2,3]

1. Department of Psychiatry, University of Oxford, Oxford, UK
2. Department of Psychology, University of Toronto Scarborough, Toronto, Canada
3. Clematis Research Empowerment Hub, Toronto, Canada
4. Denmark High School, Georgia, United States

*Correspondence to: sapolnach.prompiengchai@psych.ox.ac.uk

**Abstract**. Formative assessment is a cornerstone of effective teaching and learning, providing students with feedback to guide their learning. While there has been an exponential growth in the application of generative AI in scaling various aspects of formative assessment, ranging from automatic question generation to intelligent tutoring systems and personalized feedback, few have directly addressed the core pedagogical principles of formative assessment. Here, we critically examined how generative AI, especially large-language models (LLMs) such as ChatGPT, can support key components of formative assessment: helping students, teachers, and peers understand "where learners are going," "where learners currently are," and "how to move learners forward" in the learning process. With the rapid emergence of new prompting techniques and LLM capabilities, we also provide guiding principles for educators to effectively leverage cost-free LLMs in formative assessments while remaining grounded in pedagogical best practices. Furthermore, we reviewed the role of LLMs in generating feedback, highlighting limitations in current evaluation metrics that inadequately capture the nuances of formative feedback, such as distinguishing feedback at the task, process, and self-regulatory levels. Finally, we offer practical guidelines for educators and researchers, including concrete classroom strategies and future directions such as developing robust metrics to assess LLM-generated feedback, leveraging LLMs to overcome systemic and cultural barriers to formative assessment, and designing AI-aware assessment strategies that promote transferable skills while mitigating overreliance on LLM-generated responses. By structuring the discussion within an established formative assessment framework, this review provides a comprehensive foundation for integrating LLMs into formative assessment in a pedagogically informed manner.

# 1. Introduction

Formative assessment is a process of gathering, analyzing, and using information about students' learning to make timely instructional decisions and guide students toward their learning goals. Unlike summative assessments, which evaluate learning at the end of an instructional period (Taras, 2005), formative assessment emphasizes ongoing, real-time insights that help students, teachers, and peers understand "where learners are going," "where learners currently are," and "how to move learners forward" in their learning journey (Black & Wiliam, 2009; Wiliam, 2010). Research has shown that well-designed formative assessments enhance student engagement (Viegas et al., 2015), foster metacognitive awareness (Clark, 2012), and promote deeper conceptual understanding (C. Jacoby et al., 2014). In student-centered learning environments, formative assessment allows students to take an active role in their learning, while teachers and peers contribute valuable feedback that shapes instructional adjustments and supports student development (Connell et al., 2016; Prompiengchai et al., 2024). As educational technologies evolve, there is increasing interest in leveraging artificial intelligence (AI), particularly large language models (LLMs), to facilitate and scale formative assessment in innovative ways (Dai et al., 2024; Fütterer et al., 2023; Maity & Deroy, 2024; Yusuf et al., 2024).

Formative assessments have been conducted through various methods such as teacher observations, classroom discussions, quizzes, and self-reflections (Andersson & Palm, 2017; Jiang, 2014; Kwan, 2011; Ratnayake et al., 2024; Ruiz-Primo & Furtak, 2007). Peer reviews and group activities are also common, enabling students to learn from one another's perspectives (Phielix et al., 2009; Topping, 2005). With the advent of technology, online formative assessments have become increasingly prevalent, offering tools like real-time quizzes, interactive exercises, and digital feedback platforms (Collimore et al., 2015; Paré & Joordens, 2008). These online tools make it easier to track progress, provide immediate feedback, and adapt content to individual learners' needs. However, despite these advancements, challenges remain. For instance, providing timely and personalized feedback for large groups of students can be overwhelming for educators. Similarly, designing assessments that are both engaging and diagnostically effective requires significant expertise and effort (Mok, 2011; Prompiengchai et al., 2024; Tay & Ng, 2015). These limitations often hinder the scalability and consistent application of formative assessment, raising the question of how AI-driven technologies, particularly LLMs, can support and enhance formative assessment practices.

LLMs, such as ChatGPT, have demonstrated remarkable capabilities in generating text-based responses, personalizing instruction, and providing instant feedback (Adeshola & Adepoju, 2024). In the context of formative assessment, they have the potential to assist with automated feedback generation (Dai et al., 2023), adaptive questioning (Abolnejadian et al., 2024), scaffolding student understanding (Liao et al., 2024), and facilitating peer discussions (J. Liu et al., 2024). Despite these advancements, there is a notable lack of systematic alignment between LLM applications and the core principles of formative assessment. Consequently, it remains unclear how LLMs can comprehensively support formative assessment processes to reinforce meaningful learning outcomes. Establishing a clear framework aligning LLM applications with formative assessment principles will enable researchers and educators to systematically evaluate and enhance these technologies, ultimately leading to more effective and impactful educational practices.

Another critical challenge in LLM-enabled formative assessment is the evaluation of the quality of LLM-generated feedback. Effective formative assessment relies on different types of feedback that guide learners at multiple levels (Hattie & Timperley, 2007), including task-level feedback (addressing correctness and content-specific responses), process-level feedback (helping students refine the process used to approach the task), and self-regulation feedback (supporting independent learning and metacognitive development). Current research on LLM-generated feedback often use metrics that evaluate accuracy and coherence but does not sufficiently distinguish between these feedback processes (Chang et al., 2024; Dai et al., 2023; Hellas et al., 2023). Furthermore, there are concerns that students may over-rely on LLMs to complete formative assessment tasks rather than engaging in critical thinking and reflection (Ray, 2023; Tu et al., 2023). This raises questions about how educators can design AI-aware formative assessments that foster the pedagogical benefits of formative assessment while mitigating students' passive reliance on AI-generated responses.

Our narrative review aims to address these gaps by systematically placing LLM applications within a formative assessment and feedback framework. First, we outline the theoretical foundations of formative assessment, reviewing its core principles and the processes of formative feedback. We then examine how LLMs have been integrated into educational technology, explore prompting techniques relevant to formative assessment, and categorize their applications in the context of formative assessments. Building on this, we review how LLMs can contribute to each key dimension of formative assessment and provide guiding principles for educators to leverage them effectively. Additionally, we explore emerging techniques to enhance formative assessment practices and discuss strategies for designing LLM-driven assessments that foster deep learning and transferable skills. By synthesizing existing research and offering practical recommendations, this review provides a foundation for integrating LLMs into formative assessment in ways that are both theoretically grounded and practically viable. In doing so, we offer insights into the future of AI-enhanced assessment and discuss directions for further research, including developing robust metrics to evaluate LLM-generated feedback, leveraging LLM to address systemic barriers in formative assessments, and designing innovative formative assessment approaches in an era of increasingly powerful AI tools.

## 2. Formative Assessment, LLM, and Prompting

### 2.1 Formative Assessment: Theoretical Framework

The definition of formative assessment has evolved significantly over time. Some definitions are more restricted to assessments conducted while the learning was taking place, with the purpose of "identify[ing] specific student misconceptions and mistakes while the material is being taught" (Kahl, 2005); "recogniz[ing] and respond[ing] to student learning in order to enhance that learning" (Cowie & Bell, 1999). Black and Wiliam (2009) offered a more comprehensive definition, emphasizing that formative assessment is not bound to specific agents; it recognizes that teachers, learners, or peers can all contribute to making decisions about the next steps. Central to this definition is the focus on "next steps in instruction," broadly conceptualized as any deliberate activity aimed at fostering learning, encompassing both teaching and learning processes rather than adhering to a narrow transmissionist approach (Black & Wiliam, 2009). Importantly, the definition acknowledges that formative assessment does not necessarily require changes to planned instruction; rather, it ensures that instructional decisions are either improved or more robustly justified compared to those made without the evidence gathered through the assessment process, thereby enhancing their overall validity and effectiveness.

Formative assessment can also be conceptualized as the intersection of two dimensions: the instructional process, which includes understanding where learners are in their learning, where they need to go, and how to guide them there; and the agents of the instructional process, namely teachers, peers, and learners themselves (Broadfoot et al., 2002). This perspective, articulated by Wiliam and Thompson (2007), offers a structured approach to designing strategies tailored to each agent at each stage of the instructional process (Wiliam & Thompson, 2017). Using this approach to understanding different components of formative assessment allows us to better identify where LLMs could complement and enhance formative assessment practices. Specifically, by critically examining how LLMs could be strategically integrated into each instructional process within the formative assessment framework, we can ensure that the needs of each agent—teachers, peers, and learners—are addressed, ultimately enhancing the effectiveness and inclusivity of the assessment process.

To demonstrate how LLMs could be structured within the formative assessment framework, we translate these core principles of formative assessment into a familiar road-trip metaphor (Figure 1). The learning journey begins when teachers and students set the destination (A)—clear, student-friendly goals that define what success looks like. Like a GPS satellite giving updates on a car's position, real-time checks of understanding allow the learner (i.e., the driver) to assess where they are in their learning journey (B). At every junction, the learner makes formative adjustments or incremental decisions on which new path to take in their learning journey based on available evidence (C). The junction could be a metaphor for a formative assessment that test learners' knowledge and skills. Sometimes, students stray onto a side road or the wrong path; a good system would issue dynamic feedback that is personalized based on the student's learning journey (D). As the trip progresses, learners are prompted to monitor their own "fuel and engine" (E), building the self-regulation skills that eventually let them navigate without external prompts. Finally, on reaching the goal, the dashboard displays both arrival time and a prompt for reflection: "What will you improve for the next journey?" (F).

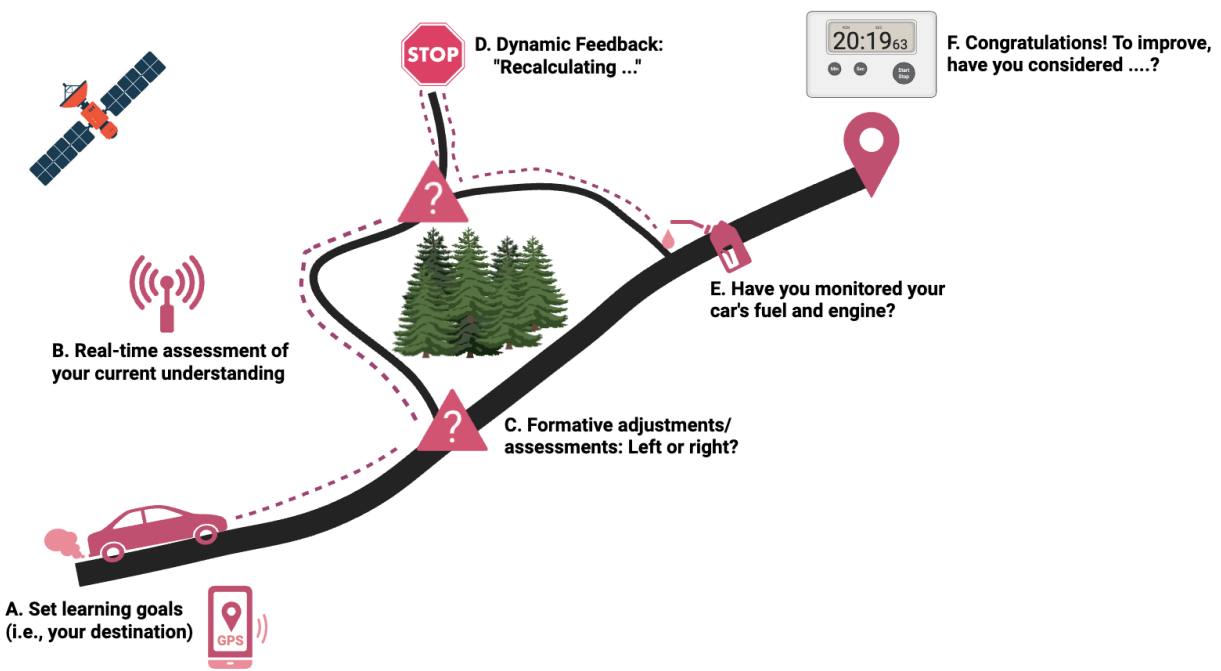

**Figure 1**. GPS-guided learning journey as a metaphor for LLM-driven formative assessment. (A) Teachers and students start by setting the destination—clear learning goals. (B) Ongoing formative assessments locate each learner (i.e., driver) on their journey. (C) The teacher and learner make formative adjustments or incremental decisions at each junction based on available evidence. (D) LLM allows for dynamic, adaptive feedback based on the student's learning journey. (E) and (F) Examples of self-regulated feedback, prompting learners to monitor and reflect on their own learning journey. The diagram emphasizes that LLM can augment, but not replace, expert pedagogy, ensuring every student reaches the destination while progressively learning to steer for themselves.

Continuing with the analogy, a large-language model functions as the in-car GPS: it can automate the generation of student-friendly goals, provide continuous location updates or frequent formative assessment, and deliver adaptive, turn-by-turn guidance at a scale that would be challenging for human teachers to achieve. Yet the driver remains the learner, with the teacher planning the route and deciding when to accept or override the AI's suggestions. By illustrating formative assessment as a technology-enhanced navigation system, the diagram underscores how LLMs can augment, but not replace, expert pedagogy, ensuring every student reaches the destination while progressively learning to steer for themselves.

## 2.2 LLM and Prompt Engineering

Large language models, such as OpenAI's ChatGPT (OpenAI et al., 2023), are advanced natural language processing (NLP) systems designed to generate human-like text by predicting the most likely sequence of words based on input (Vaswani et al., 2017). These models are trained on extensive datasets containing diverse language patterns, enabling them to learn the statistical relationships between words and phrases (Bender et al., 2021; Brown et al., 2020). Fundamentally, LLMs rely on probabilistic methods, leveraging deep learning techniques to analyze and predict text. For educators, it is important to understand that LLMs do not "comprehend" text in the way humans do but instead generate outputs based on patterns and likelihoods derived from their training data (B. Chen et al., 2024; Webson & Pavlick, 2022). This makes the input, or "prompt," critical, as it determines the model's interpretation of the task and the direction of the response it produces (Y. Lu et al., 2022; Webson & Pavlick, 2022).

In educational contexts, the role of prompting is pivotal because it acts as the bridge between the educator's instructional goals and the LLM's output. Prompts serve as the instructions or questions that guide the model in generating content aligned with specific educational needs, such as creating formative assessments, explaining concepts, or scaffolding student learning (Ouyang et al., 2022). The effectiveness of an LLM's output is largely contingent on the clarity, specificity, and structure of the prompt (B. Chen et al., 2024; L. S. Lo, 2023). A well-designed prompt can elicit relevant, accurate, and context-sensitive responses, whereas an ambiguous or overly broad prompt may lead to irrelevant or incorrect outputs (Luo et al., 2019; Song et al., 2018; M. Yang et al., 2019). For educators, we argue that prompting is, in some sense, a skill that requires careful consideration of the instructional context and the intended learning outcomes.

Basic prompting techniques that educators can easily use are zero-shot, one-shot, and few-shot prompting (Table 1). These techniques guide large language models in generating task-specific outputs. Zero-shot prompting involves providing the LLM with a task description without any examples, relying entirely on its pre-trained knowledge to generate a response (Kojima et al., 2022; P. Liu et al., 2023). This approach is efficient for straightforward tasks but may struggle with nuanced or specialized queries (Reynolds & McDonell, 2021). One-shot prompting supplements the task description with a single example, helping the model contextualize the task more effectively. Few-shot prompting expands on this by including multiple examples, providing richer context and enabling the model to perform better in more complex tasks (Logan IV et al., 2021). These techniques leverage the model's in-context learning ability, where exposure to examples within the prompt enhances its capacity to understand and respond accurately to the task at hand. However, few-shot prompts may not always be better than zero-shot prompts (J. Liu et al., 2020), and given the inherent variability in the output (Radford et al., 2019), generating responses several times can help!

**Table 1**. Illustrative Zero-Shot and One-Shot Prompt Templates for Formative-Assessment Tasks

| Task | Formative Assessment Component | Zero-shot prompting | One-shot / Few-shot prompting |
|------|-------------------------------|---------------------|-------------------------------|
| Generate Student-Friendly Learning Objectives | Clarify Learning Goals | Instruction: Rewrite the following formal learning objective in a more student-friendly way, something that is easier for students to understand and relate to.<br><br>Learning Objective: "Understand the molecular | Instruction: You are given a formal learning objective. Your task is to rewrite it in a student-friendly way, something that is easier for students to understand and relate to.<br><br>Format your response as follows: |

basis of genetic inheritance, including the structure and function of DNA, RNA, and the process of protein synthesis."

- Original Objective: "..."
- Student-Friendly Version: "..."

Example:

- Original Objective: "Understand the molecular basis of genetic inheritance, including the structure and function of DNA, RNA, and the process of protein synthesis."
- Student-Friendly Version: "By the end of this lesson, you'll be able to explain how genetic information is passed from parents to offspring. You will also learn how DNA and RNA work together to make proteins, which control how our bodies grow and function."

Now try this one:

- Original Objective: "..."

| Generate process-level feedback | Move learning forward | You are an algorithm-efficiency coach.<br><br>Context<br><br>• Student's Python script is pasted between <<< and >>>.<br><br>• The learning goal is to help the student explore *alternative strategies* for improving time / space complexity, **not** to supply finished code.<br><br>Task<br><br>1. Identify and briefly describe TWO different optimisation strategies the student could experiment with.<br><br>2. For each strategy explain **why** it may help (e.g., …).<br><br>3. Do **NOT** write or reveal complete code; speak at the conceptual, process level only.<br><br>Student submission <<<<br><br>{paste Python code here}<br><br>>>> | You are an algorithm-efficiency coach.<br><br>Context<br><br>• Student's Python script is pasted between <<< and >>>.<br><br>• The learning goal is to help the student explore *alternative strategies* for improving time / space complexity, **not** to supply finished code.<br><br>Task<br><br>1. Identify and briefly describe TWO different optimisation strategies the student could experiment with.<br><br>2. For each strategy explain **why** it may help (e.g., …).<br><br>3. Do **NOT** write or reveal complete code;<br><br>▼ EXAMPLE<br><br>Example code <<<<br><br>{ paste example Python code here }<br><br>>>> |
|---|---|---|---|

Example feedback

Return in this format:

- Strategy 1 → {one-sentence name}. *Rationale:* {2-3 sentences}.

- Strategy 2 → {one-sentence name}. *Rationale:* {2-3 sentences}.

- Strategy 1 → Use XYZ instead`. *Rationale:* reusing XYZ reduces Python-level loops ….

- Strategy 2 → Use ABC method instead *Rationale:* on typical machines, using ABC helps with DEF…

---------------------------

▼ NOW DO THE SAME FOR THIS NEW SUBMISSION

Student's Python code <<<

{paste new code here}

>>>

Return in this format:

- Strategy 1 → {one-sentence name}. *Rationale:* {2-3 sentences}.

- Strategy 2 → {one-sentence name}. *Rationale:* {2-3 sentences}.

It is very easy to generate output from LLMs. However, creating prompts that effectively aid teaching and student's learning should be informed by subject matter expertise, pedagogical insight, and a nuanced understanding of the classroom context. Moreover, educators may need to iteratively refine their prompts based on the outputs they receive, using their professional judgment to evaluate the relevance, accuracy, and utility of the generated content. Here, the educator's expertise complements the model's capabilities to create meaningful and effective learning experiences.

## 2.3 Categorizing LLM Applications in Formative Assessment

In recent years, research into the applications of LLMs in education has exploded (Fütterer et al., 2023; Kasneci et al., 2023). The integration of LLMs into formative assessment practices can be broadly categorized based on its scope and level of specialization. In our review, we have roughly categorized them into three groups: generic LLMs for broad educational applications, domain-specialized LLMs for context-specific content, and integrated LLMs within educational platforms for adaptive content generation.

### 2.3.1 Generic LLMs

First, the generic LLMs, such as ChatGPT in their standard configuration, are versatile tools capable of generating content relevant to various formative assessment contexts. Using basic prompting techniques, the models can produce diagnostic questions, summarize student responses, generate general feedback applicable across multiple subjects, and many more (C. K. Lo, 2023; Sok & Heng, 2023). While highly accessible, the quality of outputs from these models depends both on the user's ability to craft effective prompts and on the inherent capabilities and limitations of the model itself, such as its training data and reasoning capacity (B. Chen et al., 2024).

### 2.3.2 Domain-specialized LLMs

Next, LLMs can be enhanced through plug-ins or fine-tuning on datasets specific to particular domains or educational contexts (B. Chen et al., 2024; OpenAI, 2023). For example, a plugin designed for formative assessment might pull content from educational repositories to generate discipline-specific questions or feedback tailored to a particular curriculum. The use of external prompt engineering assistants like GPT plugins can streamline the generation of high-quality outputs by leveraging optimized prompting protocols that are pre-designed by the system or developers. Unlike user-created inputs or prompts, which may vary in clarity and specificity depending on the user's expertise, these optimized prompts through GPT plugins are designed to maximize the model's performance. This eliminates the need for users to manually create complex or excessively detailed prompts, making the process more accessible and reliable for those less familiar with advanced prompting techniques. However, plugins also have limitations. They are often closed-source and dependent on proprietary systems, which can restrict customization and raise concerns about transparency and data privacy (B. Chen et al., 2024). Furthermore, their utility is bound by the breadth and quality of the integrated resources, making them less flexible than standalone LLMs for unanticipated or novel tasks.

### 2.3.3 LLM-integrated Educational Technology

LLMs can also be integrated within educational platforms that support adaptive, real-time formative assessment. These systems leverage advanced prompt-engineering techniques and aim to align with pedagogical frameworks. Khan Academy's Khanmingo, powered by LLM, is one such example. The organization describes Khanmingo (https://www.khanmigo.ai/) as "an AI-powered personal tutor and teaching assistant" that "doesn't just give answers", but instead, "with limitless patience, it guides learners to find the answer themselves." By seamlessly embedding LLM capabilities into the formative assessment process, such platforms attempt to empower educators to continuously monitor and enhance student progress.

### 2.4 Scaling Formative Assessment with Accessible LLMs

Educational technologies have long played a transformative role in education, and their impact continues to expand with the rapidly growing capabilities of LLMs. Automating or facilitating processes related to formative assessment, such as AI-driven lesson planning (Tran et al., 2018), generating multiple-choice questions (X. Lu et al., 2023; Rodriguez-Torrealba et al., 2022), and providing personalized essay feedback (Botarleanu et al., 2019), traditionally required complex algorithmic processes integrated into specialized commercial applications. These systems were often costly and inaccessible, particularly for educators in low- and middle-income countries.

The advent of generic LLMs and free plugins that facilitate prompt optimization has democratized access to these technologies. With greater internet availability, educators worldwide can now leverage these tools without depending on expensive, specialized educational technologies that may need to undergo a complex procurement process (Ali et al., 2024). However, despite their potential, the effectiveness of these tools depends on how educators implement them in their classrooms. This highlights a persistent challenge: the "poverty of practice," where the capacity to use educational tools meaningfully is limited by inadequate training or understanding of their educational applications (Black & Wiliam, 1998). Furthermore, even specialized LLM algorithms or advanced optimized prompting protocols that might perform "better" than the generic LLMs have often been developed with mechanistic benchmarks rarely contextualized within formative assessment practices (Botarleanu et al., 2022; Cohn et al., 2024; Sridhar et al., 2023; Sung et al., 2019).

Thus, educators must develop a nuanced understanding of when, where, and how to integrate LLMs into their teaching practices to maximize their benefits. This review demonstrates how accessible LLM technology can enhance formative assessment practices by addressing its three core aspects: identifying students' current understanding, clarifying the learning goals they need to achieve, and providing actionable guidance to help them progress (Black & Wiliam, 2009; Wiliam, 2010). By empowering teachers to intuitively and effectively use LLMs, we aim to enable scalable formative assessment practices that allow educators to adjust their instructional strategies to meet diverse learning needs and improve educational outcomes. The following sections explore how LLMs can be embedded into these dimensions of formative assessment, creating opportunities for more responsive and impactful teaching.

## 3. Enhancing Classroom Formative Practices with LLMs

Under the pedagogical framework of formative assessment, we provide examples of how large language models can be implemented to enhance classroom formative practices, with an emphasis on the use of accessible LLMs and prompting techniques. Where relevant, we describe how domain-specialized LLMs or LLM-integrated EdTech could be developed and/or implemented to facilitate formative assessment. We also review recent advances in LLM research that are relevant to scaling formative assessment practices, identify limitations with current approaches and recommend future directions.

In some sections, we have demonstrated some prompt examples and LLM-generated outputs, but the purpose is not to rigorously evaluate or systematically test the efficiency of specific prompting approaches, as other papers and reviews have extensively addressed such evaluations. Instead, our focus lies in outlining the principles of formative assessment and identifying the components within formative assessment where LLM can be effectively integrated to create authentic assessments that inform instructional practices. The intention here is to demonstrate potential applications, make preliminary observations, and stimulate discussion on the practical implications of using LLMs in formative settings.

### 3.1 LLM in Clarifying "Where the Learner is Going"

Establishing well-defined learning objectives and ensuring that students comprehend and internalize these intentions, alongside the criteria for success, are critical to the effectiveness of formative assessment (Wiliam, 2000, 2010). When teachers clearly communicate the goals of learning activities, they create a structured pathway that facilitates purposeful teaching and learning. For students and peers, this clarity and shared understanding fosters self-regulation, enables them to monitor their progress meaningfully, and equips them to work with each other (e.g., peer feedback) that aligns with the intended learning outcomes (Nicol & Macfarlane-Dick, 2006). In other words, a shared understanding of "where the learner is going" enables effective formative assessment by allowing teachers, students, and peers to contribute to regulating learning processes. Conversely, a lack of understanding of teacher expectations often contributes significantly to underachievement (Black & Wiliam, 1998).

In practice, effectively establishing learning goals can be quite challenging. Due to time constraints, particularly in public schools, schools in developing countries, and large classrooms, teachers often limit their efforts to simply stating learning objectives that are either directly copied or minimally adapted from textbooks, curriculum guides, or state and national standards (Leahy et al., 2005). This approach is rarely effective, as the language of these objectives is often too complex or abstract for students to understand, leaving them unclear about what is expected and how to achieve success (Rust et al., 2003; Sadler, 1989; Yorke, 2003).

Recent research has explored the potential of large language models in streamlining and automating the creation of objectives, lesson plans, and curriculum design. While these advancements can assist teachers in setting learning goals, they are rarely contextualized within the framework of formative assessment. Specifically, such efforts often overlook how these tools can lay a foundation for guiding students toward achieving their goals. For instance, evaluative metrics for the success of LLMs in this domain often focus on grammatical correctness, the inclusion of action verbs, and the categorization of these verbs according to Bloom's taxonomy (Sridhar et al., 2023; Tran et al., 2018). While these benchmarks are useful for ensuring technical and structural quality, they fail to capture formative dimensions, such as how well these objectives help clarify learning expectations for students. The need to create evaluative benchmarks based on principles of formative assessment will be a recurring theme in our review.

We believe LLM can play a huge role in facilitating the implementation of several strategies that have been proposed to help students understand the intended learning goals, and we provide three examples. First, we show how generic LLMs can easily be prompted to reframe complex objectives into student-friendly language. Stiggins (2010) lists providing "a clear, understandable, student-friendly version of the learning target to the student from the beginning of the learning" as the first instructional strategy in the formative assessment. Second, LLMs can generate ideas for teachers to engineer effective classroom discussions where students co-construct learning goals. This way, students not only understand but also internalize the learning goals (Wiliam, 2010). Third, teachers can use LLMs to generate sample work of wide-ranging qualities. The use of exemplars is a powerful technique to help clarify the learning goals, criteria, and rubric assessments (Hendry et al., 2012; O'Donovan et al., 2008; Orsmond et al., 2002; To et al., 2022).

### 3.1.1 Using LLMs to Generate Student-Friendly Learning Objectives

LLMs can be idea generators through text, with the capabilities of creating high-quality diagrams, images, and video slowly catching up (Y. Liu et al., 2024; Yin et al., 2024). Thus, the first natural use of LLM is to rewrite the jargon-heavy, abstract learning objectives from the syllabus into more understandable terms. This process involves text simplification, a well-studied problem in natural language processing (Al-Thanyyan & Azmi, 2022; Niklaus et al., 2023). Recent studies have shown that off-the-shelf LLMs (e.g., ChatGPT 3.5) outperform state-of-the-art methods and are judged to be on par with human annotators across different subjects (Feng et al., 2023; Kew et al., 2023). Of note, these LLMs were generic LLMs not specifically trained for simplification.

For example, the Grade 12 Biology syllabus may state "Understand the molecular basis of genetic inheritance, including the structure and function of DNA, RNA, and the process of protein synthesis", which although is accurate, would be overly technical and abstract for many high school students. An alternative is to use (or adapt) LLM-generated materials. The teacher can simply ask ChatGPT a concise zero-shot prompt "how to make this student friendly?", and voila, we get this text: "By the end of this lesson, you'll be able to explain how genetic information is passed from parents to offspring. You will also learn how DNA and RNA work together to make proteins, which control how our bodies grow and function" (Figure 2).

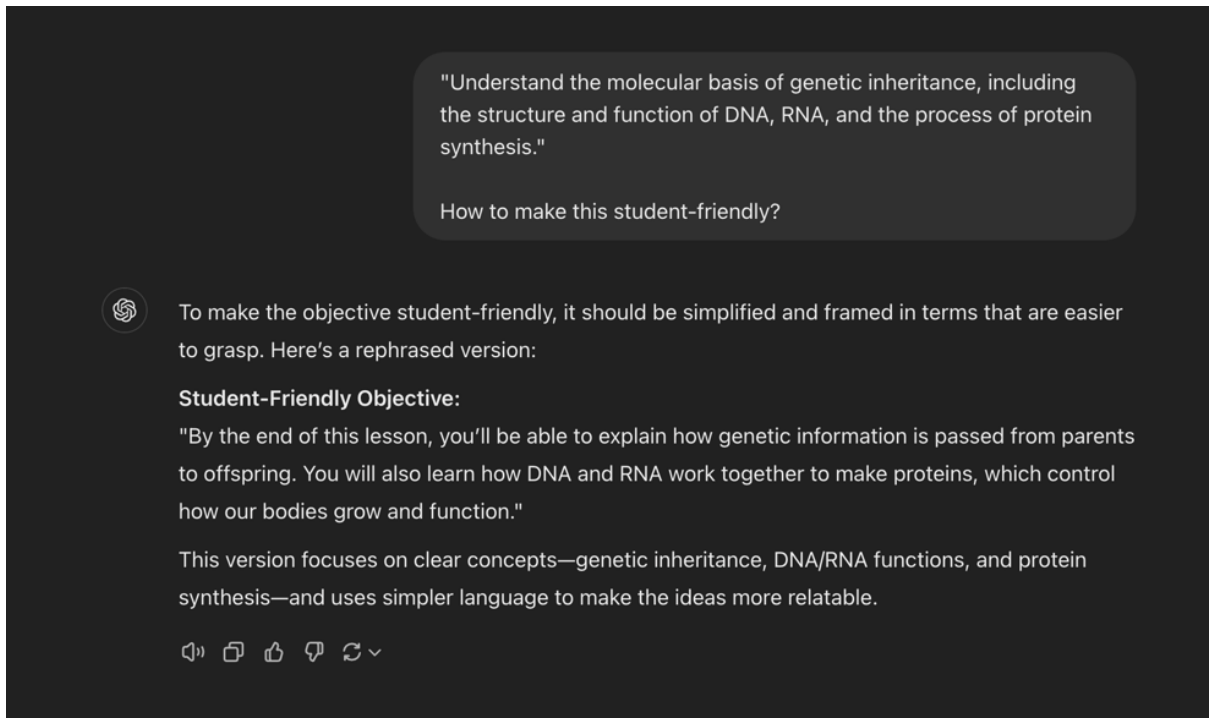

**Figure 2**. Screenshot of ChatGPT prompts to create student-friendly learning objectives.

The LLM-generated text above effectively simplified inheritance and protein synthesis. This makes sense since previous benchmark tests on GPT-3.5 show that it can respond reasonably well to zero-shot prompts (Kojima et al., 2022; Qin et al., 2023; Ye et al., 2023). And while LLMs can save time, we encourage teachers to always check the accuracy and appropriateness of the LLM-generated learning objectives for their class. For instance, teachers may point out that the goal of the learning objectives is to get students exposed to the terms that they are going to be learning, like "inheritance", "protein synthesis", "structure/function" in their learning objectives. This is where the teachers need to adapt by, for example, adding another bullet point with terms students will need to learn, or by asking ChatGPT to emphasize certain points like "Can you emphasize that students will be learning "structure" and "function" of DNA and RNA?". ChatGPT generates a new learning objective: "By the end of this lesson, you'll be able to explain how genetic information is passed from parents to offspring. You will also learn the structure and function of DNA and RNA, and how they work together to make proteins that control how our bodies grow and function."

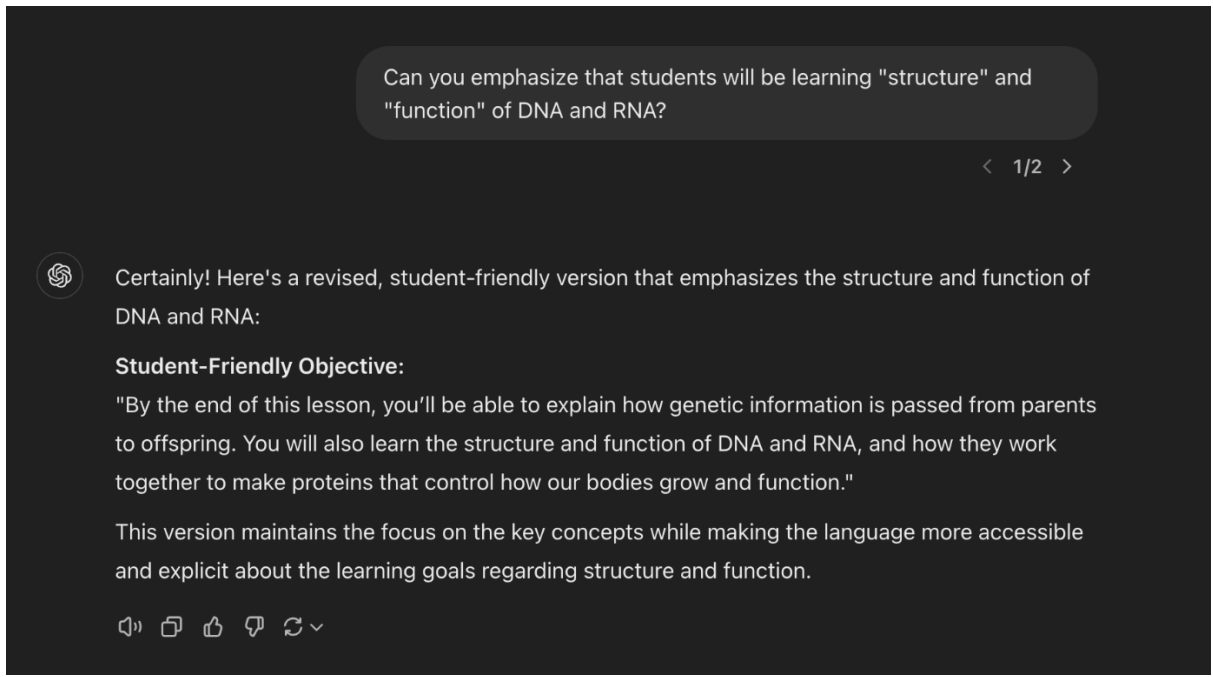

**Figure 3**. Screenshot of ChatGPT prompts to create student-friendly learning objectives, with emphasis on specific information.

Furthermore, ChatGPT can simplify large chunks of syllabus objectives, which could be distributed as handouts before class. For instance, we used the same prompt for the IB Biology Syllabus (https://chatgpt.com/share/67674fa9-e5b0-8008-a4c6-a46eae68f828), and additionally prompted ChatGPT to create a list of key important terms from the syllabus (Figure 4).

| D1.1.1—DNA replication as production of exact copies of DNA with identical base sequences |
|---|
| Students should appreciate that DNA replication is required for reproduction and for growth and tissue replacement in multicellular organisms. |

| D1.1.2—Semi-conservative nature of DNA replication and role of complementary base pairing |
|---|
| Students should understand how these processes allow a high degree of accuracy in copying base sequences. |

| D1.1.3—Role of helicase and DNA polymerase in DNA replication |
|---|
| Limit to the role of helicase in unwinding and breaking hydrogen bonds between DNA strands and the general role of DNA polymerase. |

| D1.1.4—Polymerase chain reaction and gel electrophoresis as tools for amplifying and separating DNA |
|---|
| Students should understand the use of primers, temperature changes and *Taq* polymerase in the polymerase chain reaction (PCR) and the basis of separation of DNA fragments in gel electrophoresis. |

| D1.1.5—Applications of polymerase chain reaction and gel electrophoresis |
|---|
| Students should appreciate the broad range of applications, including DNA profiling for paternity and forensic investigations. |

From the syllabus, can you create a list of key important terms from this syllabus?

Here's a list of key important terms from the syllabus:

1. **DNA Replication**
2. **Semi-conservative Replication**
3. **Complementary Base Pairing**
4. **Helicase**
5. **DNA Polymerase**
6. **Polymerase Chain Reaction (PCR)**
7. **Primers**
8. **Taq Polymerase**
9. **Temperature Changes (in PCR)**
10. **Gel Electrophoresis**
11. **DNA Profiling**
12. **Paternity Testing**
13. **Forensic Investigations**
14. **DNA Fragments**
15. **Hydrogen Bonds**

**Figure 4**. Screenshot of ChatGPT listing key important terms from the IB Biology Syllabus.

LLMs can also be used to generate learning objectives from scratch. Sridhar et al. (2023) showed that GPT-4 can automatically generate sensible learning objectives for an elaborate prompt that explains what makes a good prompt and consists of example learning objectives. However, they used template prompts carefully curated to their university guidelines on developing learning objectives, and it is unclear whether that would be compatible with other educational and cultural contexts.

Alternatively, we can rely on the basic principles of prompting to create learning objectives, such as being clear and precise with our context and instructions, and using "role-prompting", a technique that requests the model to play a specific role, such as a neuroscience professor teaching fourth-year undergraduate course. For example, we prompted this input below, asking ChatGPT 4o-mini to generate learning objectives for this very specific module. Important contexts such as the knowledge level (e.g., students have already learned the basics of fMRI experimental design and data analysis) are necessary. As we can see, ChatGPT did reasonably well in generating learning objectives (Figure 5). But as with other outputs, adjustments are probably warranted:

> *"I want to teach my class of undergraduate students in my Introduction to Functional Magnetic Resonance Imaging class to write a method section for an fMRI paper. They have already learned about the basics of fMRI experimental design and data analysis. Write student-friendly learning objectives for today's class."*

**Learning Objectives: Writing the Methods Section for an fMRI Paper**

By the end of today's class, you should be able to:

1. **Understand the structure of the methods section**: Recognize the key components of a methods section in an fMRI paper, including study design, participants, materials, data acquisition, and data analysis.

2. **Describe the experimental design**: Be able to clearly explain how participants were involved in the study (e.g., task design, conditions, counterbalancing) and how the fMRI experiment was structured (e.g., trial sequence, timing).

3. **Detail participant information**: Accurately describe participant characteristics (e.g., number, age range, inclusion/exclusion criteria) in a way that is reproducible and transparent.

4. **Outline data acquisition methods**: Clearly describe the fMRI scanning protocol, including the type of scanner, imaging sequence parameters (e.g., TR, TE, voxel size), and any relevant acquisition details.

5. **Explain data preprocessing steps**: Identify and summarize common preprocessing techniques (e.g., motion correction, normalization, smoothing) and justify their use in an accessible way.

6. **Report data analysis strategies**: Write about statistical analyses used, such as GLM (General Linear Model), contrasts, and corrections for multiple comparisons, ensuring clarity and transparency in explaining the rationale for each method.

7. **Use clear and precise language**: Communicate complex technical details in a clear, concise, and accessible way, so that readers without prior knowledge of the study can understand the methods.

**Figure 5.** Screenshot of ChatGPT generating learning objectives for a fourth-year undergraduate neuroimaging (functional MRI) course.

Since large language models generate text based on probabilistic predictions of word sequences, variations in the output are inevitable, even for the same prompt, as the model may produce different responses depending on subtle variations in its internal processing or interpretation of the input (Holtzman et al., 2020). As a result, teachers may encounter variations in the output that may not meet expectations or be suitable for the given prompt. Additionally, since ChatGPT has memory capabilities (OpenAI, 2024), the output can be influenced by prior interactions, further contributing to variations that may not always align with the teacher's expectations or the intended response for a given prompt.

Using the same fMRI example, we have rerun this prompt using one of the authors' personal ChatGPT 4o-mini (https://chatgpt.com/share/676761c1-3a08-8008-883c-f70c9823f7f8). For example, the professor may not favor this variation: "Students will be able to **describe how experimental design elements** (e.g., task types, stimuli, and control conditions) **are reflected in the methods section,**" because the objective of the class is to enable students to articulate experimental design within the methods section, rather than simply describe their reflections there (Figure 6)

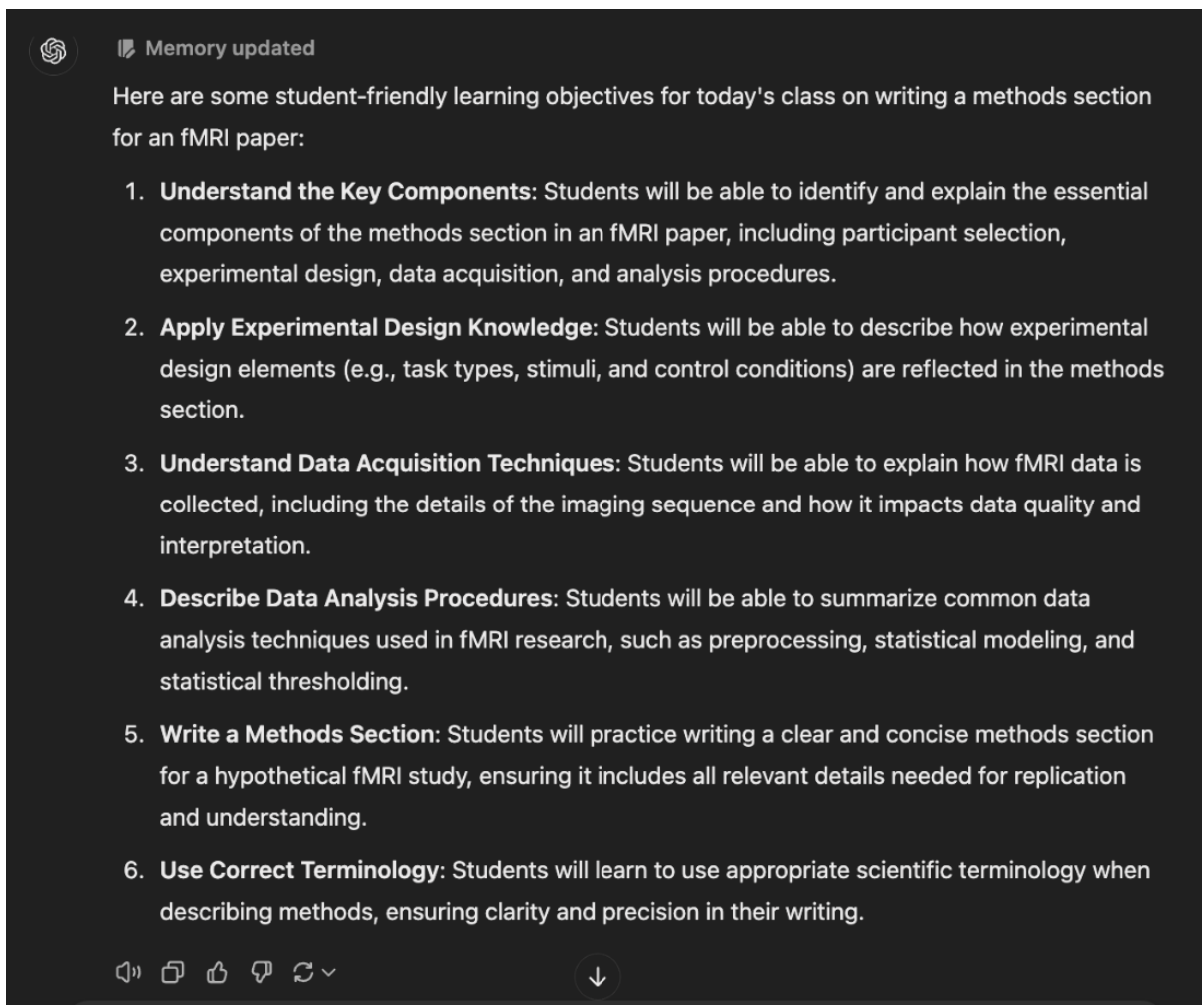

**Figure 6.** Screenshot of the author's personal ChatGPT generating learning objectives for a fourth-year undergraduate neuroimaging (functional MRI) course

Through these examples, we simply want to point out the possibility that 1) educators can use generic LLMs to generate student-friendly learning objectives, 2) educators can make the existing curriculum syllabus more student-friendly and fine-tune to specific context through clear and precise prompts, and 3) not all variations produce equally good output. Indeed, there are many informal resources (e.g., [UofMich](#)) that have prompt templates for creating learning objectives for their courses, and the field of LLM is rapidly evolving. Complementing the benchmark tests and breakthroughs in prompt engineering, future research should foster collaboration between computer scientists and educators to link pedagogical theory with practical applications and create repositories of prompt templates for lesson objective generation that have clear pedagogical explanations on their purpose, appropriate contexts for use, and strategies to maximize their effectiveness.

### 3.1.2 LLM and Co-Construction of Success Criteria

Communicating learning goals does not have to be one-way, from teacher to student. It can be a reciprocal and collaborative effort. A challenge to this is to come up with questions that will strike a meaningful discussion in co-constructing a learning goal. A large-language model can facilitate teachers in generating such questions. Although it is unclear how the LLM compares to educators in stimulating class discussions geared towards co-constructing learning goals, we show an example of how the general-purpose LLM can provide reasonable instructions for educators.

For instance, a sample prompt for the fMRI professor could be "I want to teach my class of undergraduate students in my Introduction to Functional Magnetic Resonance Imaging class to write a method section for an fMRI paper. They have already learned about the basics of fMRI experimental design and data analysis. Instead of reading out learning objectives, I want to co-construct learning objectives with students in my class. How do I do this?"

Here, ChatGPT gives a step-by-step instruction and provides the rationale for each step. For example, ChatGPT explains how the professor should first explain how the writing method section requires "both a good understanding of the experimental design and the data analysis techniques they've already learned." It suggests that the professor asks students "open-ended questions to stimulate thinking and engage the class":

- "What is the purpose of the methods section in a research paper?" to "help students realize the importance of clear and precise description of methods so others can replicate the study"
- "What components should a methods section include when writing about fMRI experiment?" to "help them focus on critical elements like experimental design, participants, stimuli, fMRI acquisition, data analysis procedures, etc."

ChatGPT also gave suggestions to ask students to translate ideas to objectives: e.g., "I can describe the fMRI experimental design clearly and precisely." (Appendix A)

The examples above demonstrate how LLMs can be employed in downstream educational tasks, such as generating ideas for educators to co-construct success criteria with students. Yet, it might be challenging for teachers to visualize how such discussion may play out in a real classroom. With the emerging research on LLM-empowered agents and human-scenario simulations (Gao et al., 2024; Park et al., 2023; Zhang et al., 2024), teachers might be able to practice co-constructing success criteria with LLM-powered "virtual students" before implementing them in real classrooms in the near future. However, developing an interactive teacher-training platform using LLM-simulated students is still in its infancy, with recent attempts to create a chat-based teacher-training tool simulating office hours (Markel et al., 2023), and a GPT-powered 3D virtual reality platform to enhance pre-service teachers' problem-solving skills in the classroom (U. Lee et al., 2023).

### 3.1.3 Using LLMs to Generate Sample Work

A common technique for ensuring that learning goals are transparent at the beginning of a class is to engage students in discussions about sample essays or other work, analyzing their quality and identifying elements that meet or fall short of the learning criteria. This approach helps students better understand the expectations and standards for their own work (Hendry et al., 2012; O'Donovan et al., 2008; Orsmond et al., 2002; To et al., 2022). However, selecting or creating sample work that effectively illustrates a range of qualities can be challenging and time-consuming for educators.

LLMs can assist educators by generating sample work tailored to varying levels of quality, enabling teachers to focus more on facilitating discussions than on creating examples, but there remains a notable gap in the literature concerning their efficacy for formative assessment purposes. Related work has shown that ChatGPT-generated argumentative essays are better than those of students (Herbold et al., 2023), and the discussion on LLM-generated writing content focusses on the implications of academic integrity (Stokel-Walker, 2022). However, LLM-generated essays still exhibit certain linguistic characteristics that are different than humans (Guo et al., 2023), and lag behind in generating emotional responses (Zhao et al., 2023). Future work should address how to tap the generic LLMs or the fine-tuned, domain-specialized LLMs to generate sample work across disciplines and educational contexts at the educator's command. Educators can also exploit the limitations of LLM-generated "samples" and ask students to critically think about how to improve the essay, which would help students internalize the learning goals (e.g., the rubrics).

Meanwhile, educators can take advantage of the generic LLMs and produce reasonable sample work. For instance, let's take an example of a professor teaching an fMRI class to undergraduate students who might want to help them develop skills in writing a clear and effective methods section for an fMRI research paper. The professor decides to begin a session by having the class critique and improve sample methods sections. To illustrate common pitfalls and strengths, the professor uses an LLM to generate three examples: one with incomplete reporting of key design elements (e.g., vague description about experimental designs), another with unnecessary details about the researchers clicking the buttons in the software graphical user interface, and a third that provides a comprehensive and well-structured description.

By prompting the LLM with a specific context (e.g., teaching undergraduate students to write an fMRI method section) and instructions like "I want to begin a session by having the class critique and improve sample methods sections. For this method section below describing experimental task, can you modify it to make it a very low quality one?", the professor can quickly generate targeted examples that highlight specific issues. The example we have shown is of Godlewska et al. (2012). As you can see, ChatGPT generated a reasonable low-quality version of the method section.

**Table 2**. LLM-generated sample work

| Original fMRI experimental task description (Godlewska et al., 2012) | ChatGPT-generated low-quality version of the fMRI experimental task description |
| --- | --- |
| **Functional MRI experimental task** | **Functional MRI experimental task** |
| During fMRI scanning, participants completed a simple gender discrimination task involving the rapid presentation of fearful and happy facial expressions. In this task, nine 30-s blocks of a baseline fixation cross (condition A) were interleaved with eight 30-s blocks of the emotional task [four blocks of fear (condition B) and four blocks of happy (condition C)]. During each emotional block, participants viewed 10 emotional faces (five female) all derived from a standard set of pictures of facial affect | In this study, participants did a simple gender task during fMRI scanning. They looked at faces showing either fear or happiness. The task had 9 blocks with some fixation cross and 8 blocks with emotional faces. The fixation cross was shown for 30 seconds and the emotional faces were shown in 30-second blocks too. In the emotional blocks, there were 10 faces, half of them were women. The faces were shown quickly (100 ms each) and participants had to press a button to say if the face was male or female. There were |

(Matsumoto & Ekman, Reference Matsumoto and Ekman1988). Each face was presented for 100 ms and subjects were asked to report the gender of the face via a MRI compatible key pad. To ensure participants remained focused during the task within block inter-stimulus intervals (ISI) ranged between 2500 and 2900 ms (mean ISI = 2900 ms). To reduce potential carry-over effects, cycles of alternation between conditions were counterbalanced across subjects. Thus, during the course of the 8.5-min experiment, half of the subjects completed the following order: ABACABACABACABACA, the remaining subjects ACABACABACABACABA. Stimuli were presented on a personal computer using E-Prime (version 1.0; Psychology Software Tools Inc., USA) and a cloned projection displayed to participants on an opaque screen located at the head of the scanner bore, which subjects viewed using angled mirrors. Subject responses were made via an MRI-compatible keypad. Stimulus presentation/subject button presses were registered and time-locked to fMRI data using E-Prime. Both accuracy (correct gender discrimination) and reaction times were recorded. Immediately before scanning, all subjects received training with another set of stimuli to ensure they fully understood the requirements of the task. Behavioural data were analysed using a repeated measures analysis of variance (ANOVA) model with group as the between-subjects factor and valence as breaks between faces, but the exact timing varied. The order of the blocks changed between subjects, so some subjects did fear and happy faces first, while others did happy and fear faces first. The stimuli were presented using a computer and shown on a screen that participants saw through mirrors. Responses were recorded using a keypad and the computer program tracked the timing of the button presses. Accuracy and reaction time were recorded. Before the scan, participants were trained on a different set of faces so they would know what to do. Data analysis was done using SPSS.

## Problems with this version:

1. Lack of specificity: The method section is vague about key experimental details, such as the exact nature of the stimuli, the timing of the blocks, and how the counterbalancing was handled.
2. Ambiguity in task description: "Simple gender task" and "faces showing either fear or happiness" are very vague and do not clearly specify how these stimuli were selected or categorized.
3. Inconsistent details about timing: The ISI timing is mentioned, but the variability is not clearly stated or explained. There is also a statement that the ISI "varied" but no concrete values are given.
4. Unclear counterbalancing: The order of conditions

the within-subjects factor implemented in SPSS v. 15 (SPSS Inc., USA).

(ABACABAC...) is presented in an unclear way, and the explanation of counterbalancing is minimal and confusing.

5. Methodological clarity: The explanation of how stimuli were presented (E-Prime and the projection screen) is not detailed enough, and it's unclear how the response tracking works.

6. No mention of pre-scanning or post-scanning procedures: The only pre-scan detail is the training, but there's no information on the consent process, or any behavioral assessments beyond reaction time and accuracy.

7. Lack of statistical detail: The statistical analysis is mentioned briefly, but the exact nature of the analysis (e.g., how group and valence factors were incorporated) is not sufficiently clear.

This would serve as a great starting point for a critique and class discussion on how to improve these issues!

Or similarly, when prompted to add unnecessary details about the researchers clicking the buttons in the FSL graphical user interface for describing fMRI data pre-processing, ChatGPT went overboard in adding the unnecessary details, driving the point home that overly specific descriptions can detract from the clarity and usefulness of a scientific methods section.

"Within the FEAT setup window, the researchers clicked the "Pre-statistics" tab. With the mouse pointer hovering over the tab for a brief moment, they selected the "Slice Timing Correction" option, ensuring the correct slice acquisition order was specified from the drop-down menu." (Appendix B)

Additionally, when high-quality exemplars are generated solely based on instructions provided to an LLM in a zero-shot approach (i.e., without supplying the model with example exemplars), educators are likely to identify inherent flaws or limitations in the LLM-generated outputs. This provides a valuable opportunity for integrating these outputs into pedagogical practices. For instance, students can be tasked with critically evaluating and differentiating between LLM-generated exemplars and authentic, high-quality versions of the same work. Such exercises not only address the growing concern of student over-reliance on tools like ChatGPT but also encourage the development of critical thinking skills. In the broader context of formative assessment, leveraging LLMs in generating sample work can help students directly engage with the learning goals while simultaneously addressing ethical and academic integrity concerns associated with AI-assisted learning tools.

## 3.2 LLM in Understanding "Where the Learner is"

To establish how LLMs can be used to formatively assess "where the learner is" in their learning trajectory, it is important to clarify what makes an assessment formative. An assessment that merely identifies a learner's current level of understanding is not formative (Black & Wiliam, 2009; Wiliam, 2010). The assessment becomes formative when the information from such assessment can be utilized to enhance the learning process or to inform teaching strategies aimed at achieving specific learning objectives. For example, feedback that simply prompts a student to "work harder" without clarifying actionable steps or providing a framework for improvement is not formative.

### 3.2.1 Cycle Lengths in Formative Assessments

An important concept in formative assessment is cycle lengths (Wiliam, 2010). Formative assessment operates over different temporal cycles, ranging from long-term evaluations to immediate feedback within a single lesson.

- Long-cycle assessments focus on periods such as semesters or academic years, with their results used to plan or adjust curricula and instructional strategies over extended timeframes (4 weeks to 1 year).
- Medium-cycle assessments address learning within specific instructional units or weeks (1 to 4 weeks).

- Short-cycle assessments provide rapid feedback within or between lessons, with lengths varying from days to minutes (5 seconds to 48 hours).

For instance, a long-cycle assessment might involve analyzing trends in a student's performance across semesters to identify areas requiring targeted intervention. Perhaps, the majority students performed poorly on the geometry-related questions for their Grade 9 Mathematics Final Examination, and the school subsequently asked the teachers to attend professional development workshops in teaching geometry (Wiliam, 2010). The teachers then changed their pedagogical techniques in the subsequent years, and the students improved their performance in geometry-related questions. Short-cycle assessments, by contrast, might include strategies like quick checks for understanding during a lesson or digital quizzes completed overnight. These assessments are critical in addressing immediate misconceptions and tailoring instruction in real time.

Some might argue that the long-cycle assessment example is summative (Popham, 2006), and that an assessment is formative because it can immediately be used to improve the instructional process and learning (Shepard, 2017). However, we adopt the view of (Wiliam, 2010) that the key question should not be, "Was this assessment formative?" but rather, "How has the use of this assessment facilitated sustainable improvements in learning?" (Kluger and DeNisi, 1996). Thus, regardless of the cycle lengths, the goal of the LLM in understanding "where the learner is" within the context of formative assessment is to enhance the assessment of student's current learning such that its result can inform subsequent instructional processes and facilitate sustainable improvements in learning.

Building on this perspective, short-cycle assessments stand out as real-time evaluations of students' understanding, offering immediate insights into where learners are in their educational journey. These assessments are indispensable in day-to-day classroom practice, enabling teachers to adapt instruction promptly to address learning needs as they arise. While this subsequent sub-section emphasizes the potential of LLMs in designing short-cycle assessments due to their accessibility and ease of use, it is essential to recognize that medium- and long-cycle assessments can also serve evaluative purposes. Techniques such as peer assessment or extended feedback-driven tasks are invaluable for fostering sustainable learning improvements (Topping, 2009). These approaches, particularly their role in guiding learners toward their goals, will be explored in subsequent sections on leveraging LLMs to support the broader formative assessment process.

### 3.2.2 LLM and Automatic Question Generation

Implementing formative assessments to test student's understanding becomes challenging particularly when teachers want to implement these assessments frequently (Madri & Meruva, 2023). A critical difficulty lies in developing materials that effectively elicit evidence of learning across diverse modalities (Lai et al., 2016; Van der Linden & Glas, 2010). Although this process is time-intensive and requires careful consideration to ensure alignment with varying grade levels, subjects, curricula, and the practical realities of teaching in large classrooms, LLMs can facilitate this process and may be able to save valuable time when used appropriately.

Since LLMs have been pre-trained on huge amounts of data, they are becoming a generalist model that can perform language tasks with no training (Qin et al., 2023). LLMs are now being used in automated question generation and assessment (Maity et al., 2023; Maity & Deroy, 2024; Mulla & Gharpure, 2023), including multiple-choice questions (McNichols et al., 2024; Moore et al., 2024). Effective formative MCQs require high-quality distractors that can check students' misconceptions about a certain topic. At the moment, generic LLM cannot consistently produce MCQs with distractors that satisfy the human experts or benchmarks (Bitew et al., 2023; J. Lee et al., 2024), and fine-tuned LLMs are also restricted to generating MCQs for certain question types and disciplines (Li et al., 2024; McNichols et al., 2024). Previous research also used different benchmarks for evaluating the quality of MCQs such as human educators (Biancini et al., 2024; Bitew et al., 2023) and automatic quantitative metrics (de-Fitero-Dominguez et al., 2024).

Yet, we believe that teachers can already start using prompts that were previously experimented using generic LLMs to generate MCQs. For instance, Biancini et al. (2024) crafted the following prompt template for MCQ generation (Figure 7). Although they don't mention the details for how they arrived at this prompt, we can see that they used basic principles of prompt engineering. First, the prompt consists of a role "You are an assistant that doesn't make mistakes". Then, the prompt provides clear and precise instructions with context, which is to create high school level quiz based on the provided text. Additional text is typed at the bottom { text } and could be the content material taught in class.

```
You are an assistant that doesn't make mistakes.
If a reference model is presented to you, you follow it perfectly
without making errors.

Create a high school - level quiz based on the provided text.
You must strictly adhere to the following format without any errors:
> [ Insert the question ]
a ) [ Option A ]
b ) [ Option B ]
c ) [ Option C ]
d ) [ Option D ]
* Correct Answer: [Insert the letter corresponding to the correct
answer for example : 'a)']
* Source: [Write the exact line or passage from the provided text
where the information for this question can be found.]

Please note that you are allowed to modify only the parts within
brackets ([...]) in the format provided.
Ensure that all four options are distinct.
When mentioning a date, please make sure to specify the year.
The text is : { text }
```

**Figure 7**. LLM prompt template for MCQ generation from Biancini et al. (2024).

However, even without access to test-and-tried prompt templates, the teachers can customize their own prompts to generate MCQs. For instance, consider the prompt below by a hypothetical Grade 12 biology teacher preparing to teach protein synthesis. Importantly, the prompt clearly specifies the desired level of difficulty, the scope of content, and/or the nature of the question (e.g., conceptual or factual). They may need to clarify what the students would not know.

> "Generate five multiple-choice questions for Grade 12 biology students to assess their foundational knowledge of DNA. The questions should focus on the structure and function of DNA, the role of nucleotides, and the difference between DNA and proteins. The questions should be factual. Include one correct answer and three plausible distractors for each question. These questions should not test and require knowledge of transcription, translation, and processes related to protein synthesis. Provide rationales for each distractor."

Designing effective multiple-choice questions (MCQs) tailored to individual classroom contexts, including students' prerequisite knowledge and instructional needs, poses significant challenges when using large language models. While LLMs like ChatGPT are capable of generating content-rich questions, asking them to account for nuanced individual-level contexts—such as students' prior knowledge, cognitive readiness, and classroom dynamics—remains a complex task.

Furthermore, the first attempt at prompting may often yield suboptimal results, underscoring the need to iteratively experiment with prompting. Even with seemingly refined prompts from the teacher's perspectives, the model may generate inappropriate questions due to its reliance on probabilistic patterns in its training data rather than a full understanding of the classroom-specific context.

For example, consider the following prompt tested with ChatGPT 4o-mini (the free unlimited version available as of December 2023):

> "I will be teaching Grade 12 Biology students today about protein synthesis. However, they would need to understand the basic knowledge of the structure and function of DNA, differences between DNA and RNA, and the difference between DNA and protein to understand my class. Assume they do not know anything about protein synthesis yet. I want to conduct an MCQ (formative assessment) test before my protein synthesis class to ensure that they have this knowledge. Can you produce MCQs with good distractors? Explain the rationale for each distractor."

Although the prompt appears to make the scope of knowledge and context clear, the model consistently suggested inappropriate questions, such as:

> "MCQ 4: What is the main role of messenger RNA (mRNA) in protein synthesis?"

This question is inappropriate because it assumes that students already have knowledge about protein synthesis, contradicting the prompt's directive to focus on prerequisite concepts like DNA structure and function. From a computational perspective, this misalignment likely arises from the model's tendency to prioritize commonly associated patterns in biological education (e.g., RNA and protein synthesis) over strictly adhering to the nuanced constraints of the prompt. This reflects the model's reliance on statistical associations from its training data, rather than its ability to infer and apply specific contextual requirements.

Despite these limitations, LLMs hold immense potential for automatic question generation. Their ability to quickly produce a variety of questions offers educators valuable starting points, provided that teachers iteratively refine and evaluate the outputs. This iterative process of prompting mirrors the broader learning journey in which educators improve their understanding of effective LLM use in classroom contexts.

## 3.3 LLM, Feedback, and Moving Learners Forward

Feedback resulting from formative assessment helps bridge the gap between existing knowledge (i.e., where the learners are) and intended learning goals (i.e., where the learners are going) to move learning forward. Feedback serves a dual formative purpose: for teachers, feedback offers valuable insights into student progress and misconceptions, enabling them to adjust their instructional strategies to better align with learners' needs and optimize the effectiveness of their teaching. For students, it provides critical information to reflect on their current understanding and refine their learning strategies, fostering self-regulation and deeper engagement with the material.

However, there has been the need for scaling customized feedback that helped personalize learning. One prominent area of progress involves intelligent tutoring systems (ITS). ITS is a computer-based learning system that integrates AI to imitate human tutors, with the goal of providing personalized instruction and feedback (Carbonell, 1970). Before LLM, the feedback generation relied on experts to create a set of rules or solutions that were then used to program feedback for all possible answers (Aleven et al., 2006, 2009; Stamper et al., 2024). However, it is very time consuming and expensive, and prone to human errors since the system depends on "experts" (Stamper et al., 2024). Instead of using experts, we can use algorithms and computational models to analyze patterns in student submissions or expert solutions, which can be used to create a set of data-driven rules for generating feedback (Conati et al., 1997; Di Eugenio et al., 2005; Stamper et al., 2024). Unfortunately, these data-driven rules still depend on the quality of training data, and the generated feedback is also dependent on that data, making the feedback less adaptable.

With LLMs, we can generate human-like feedback without training data, overcoming limitations faced by data-driven and expert-driven rules guiding feedback generation. LLMs have been explored to score and provide feedback for essays (Kortemeyer, 2023; Matelsky et al., 2023; Mizumoto & Eguchi, 2023, 2023; Yoon, 2023), integrated to gamified ITS (Cao, 2023), code review and feedback (Finnie-Ansley et al., 2022; Nguyen et al., 2024; Nguyen & Allan, 2024), but still face limitations such as inaccurate and inconsistent feedback (Balse et al., 2023). Despite the potential of LLM-generated feedback, very limited works exploring LLM-generated feedback are grounded in principles of learning and the pedagogical design of feedback (Stamper et al., 2024).

In particular, future work should specifically address how LLMs and prompt engineering can facilitate different levels of feedback proposed by Hattie and Timperley (2007). Each level addresses a specific dimension of the learning process: 1) the task, 2) the process, 3) self-regulation, and 4) the self as a person (Table 3). Most LLM studies treat feedback as a unitary component that is used to improve student learning. For instance, Nguyen et al. (2024) evaluated the LLM-generated feedback based on whether the feedback captures errors (i.e., task-level feedback). Similarly, Koutcheme et al. (2024) qualitatively evaluated the feedback based on completeness (whether it identified all issues), perceptivity (identifies at least one real issue), and selectivity (does not generate non-existent issues), all of which are specific criteria at the task-level. Below, we review recent advances in the application of LLMs for each type of feedback.

**Table 3**. Description and examples for different levels of feedback proposed by Hattie and Timperley (2007).

| Types of Feedback | Description | Example |
| --- | --- | --- |
| Task | Provides information about the specific task itself (e.g., correctness of a student's work) | Your answer needs to include more detailed explanation about protein synthesis. |
| Process | Provides insights into the strategies and methods used to approach the task | You may be able to reach the correct conclusion if you break the problem into smaller steps. |
| Self-regulation | Encourages students to monitor, direct, and reflect on their own learning and performance | You already know how a "for-loop" work in Python. Check whether you could use the loops to make your code more concise. |
| Self as a person | Provides feedback that focuses on the individual's personal qualities or identity | You are a hard-working student! |

### 3.3.1 Task-Level Feedback

Task-level feedback focuses on providing information about the correctness or quality of a specific task, such as identifying errors in an assignment or clarifying misconceptions.

Feedback about the task is the most common type of feedback in a classroom (Airasian, 1997). At the basic level, LLM can identify errors and give suggestions for structured tasks like computer programming (Balse et al., 2023; Leinonen et al., 2023). However, task feedback is most effective when addressing faulty interpretations rather than merely pointing out errors or gaps in knowledge (Hattie & Timperley, 2007). For instance, highlighting why a student might have misunderstood a concept can lead to deeper learning compared to simply providing the explanation to the correct answer. Most studies focus on the latter (Dai et al., 2023; Koutcheme et al., 2024; M. Liu & M'Hiri, 2024; Nguyen & Allan, 2024; Riazi & Rooshenas, 2024; Sarsa et al., 2022) and we argue it is because of their focus on feedback metrics like usefulness, clarity, or accuracy that do not stratify the nuances of task-level feedback.

### 3.3.2 Process-Level Feedback

Process-level feedback goes beyond the task to offer insights into the strategies and methods used to approach the task (Hattie & Timperley, 2007). For instance, the teacher might suggest the learner to think about an alternative strategy to solve a particular math problem set. Nguyen and Alan (2024) engineered prompts to provide formative code feedback for computer science students completing complex programming tasks. Although they did not show the exact prompt, the example outputs had elements of process-level feedback as "hints" or alternative steps to approaching the coding problem.

> "Hint": "Java doesn't support slicing as Python does. However, you can achieve the same effect using recursion with indices. In this case, the middle index serves as a marker to divide the array. Remember to account for both halves of the array when the middle element is 1." (Nguyen & Allan, 2024)

### 3.3.3 Feedback and Self-Regulated Learning

Formative assessment encapsulates self-regulated learning (Clark, 2012). Similar to formative assessment, a self-regulated learner will plan their goals and learning strategies, execute their plan while monitoring their actions, and self-reflect and adapt based on feedback (Zimmerman, 2000). Self-regulation feedback can target the learner's ability to monitor, direct, and evaluate their own learning (Hattie & Timperley, 2007), empowering students to become self-regulated learners and activating them to be owners of their own learning (Wiliam, 2010). It goes beyond process-level feedback by encouraging students to self-assess their knowledge, strategies, or self-beliefs, which may prompt the students to modify their strategies and achieve their intended learning goals. For instance, the teacher could give self-regulation feedback as follows: "You already know the key features of the opening of an argument. Check to see whether you have incorporated them in your first paragraph" (Hattie & Timperley, 2007).

However, the efficacy of feedback about self-regulation depends on the student's capacity to be an effective self-regulated learner. A self-regulated learner can judge when to seek help and what resources they need to succeed. They self-assess, create internal feedback, have the metacognitive skills, and take the effort to reflect about the feedback they receive. Thus, a critical component of formative assessment and self-regulation feedback is to encourage students to become self-regulated learners. There are many pedagogical techniques aimed at promoting skills related to self-regulated learning such as self-assessment, reflection exercises (Butzlaff et al., 2018; P. Chen et al., 2017; Edlund, 2020; Langdon et al., 2019; Ratnayake et al., 2024), peer assessment, and peer feedback (Collimore et al., 2015; Noroozi & De Wever, 2023; Prompiengchai et al., 2024; Topping, 2009). Self- and peer-assessment are often administered within the cloud-based platform, which provides a digital infrastructure that can automate several administrative aspects of such assessments (Collimore et al., 2015; Darvishi et al., 2022; Paré & Joordens, 2008; Prompiengchai et al., 2024).

Measuring the efficacy of LLMs in providing high-quality self-regulated feedback is limited, but there are examples where LLMs have been integrated into peer feedback exercises that potentially provide self-regulated feedback. For instance, PeerGPT utilized GPT-3.5 as conversational agents to foster peer conversations within children's collaborative, design-based learning setting (J. Liu et al., 2024). Using thematic analysis, they found that children frequently engaged with GPT-based agents for advice, and the GPT-based agents were able to initiate, facilitate, structure, and extend conversations. However, as the workshop progresses, GPT-based agents cannot perceive much of the children's non-verbal communication, resulting in the children ignoring the agent's suggestions. This suggests that, particularly for children, LLMs should be integrated alongside human facilitators who can interpret social cues, contextualize feedback, and ensure that AI-generated suggestions remain relevant and engaging in dynamic learning environments.

While limited literature explicitly investigates the use of LLM in generating self-regulated feedback, LLM-based feedback in general may improve learners' self-regulated learning, leading to improved academic outcomes (Sun et al., 2025). However, it is unclear whether general LLM-assisted learning environments can improve learners' self-efficacy (Yan et al., 2025). In particular, more studies are needed to examine how prompt engineering can be leveraged to refine the linguistic features of LLM-generated feedback in ways that enhance self-efficacy within educational settings. Understanding how specific phrasing, tone, and scaffolding strategies influence learners' confidence and motivation could inform best practices for integrating LLMs into formative assessment frameworks.

# 4. Practical Guidelines and Future Directions

Building on the theoretical and empirical foundations discussed in previous sections, this section provides practical guidelines for educators and researchers to effectively integrate large language models into formative assessment. While LLMs hold promise for enhancing feedback, fostering self-regulated learning, and personalizing instruction, their implementation requires careful consideration to align with FA principles. In this section, we summarize our review and aim to bridge research insights with practical strategies for researchers and educators interested in developing and integrating LLMs in formative assessments.

## 4.1 Integrating LLMs into Formative Assessment Framework

When designing LLM-assisted pedagogical tools aimed at improving student's learning in a formative manner, educators and researchers should intentionally specify each instructional process within the formative assessment framework that they are targeting. There are mainly three instructional processes within formative assessment that LLM can be integrated:

- Is LLM used to clarify "where the learner is going" (i.e., their learning goals)?
- Is LLM used to help students understand "where the learners are" in their learning?
- Is LLM used to help "move learning forward" by bridging the gap between existing knowledge (i.e., where the learners are) and intended learning goals (i.e., where the learners are going)?

Furthermore, it is very important that educators understand how each of the component fit into the larger purpose of formative assessment. Despite the varying definitions of formative assessment, the key is that each component of formative assessment focuses on informing the "next steps in instruction", a process in which teachers, learners, or peers can all contribute. For each implementation of LLM, it may be a good exercise for educators to write two statements, one linking LLM to one of the three formative assessment components, and another linking to the big-picture pedagogical outcome. We summarized the application of LLMs for formative assessments (Table 4).

**Table 4**. Summary of LLM-driven formative assessment and sample prompts

| LLM-driven Task | Purpose | Illustrative Prompt | Big-picture Rationale |
|---|---|---|---|
| Rewrite formal objectives in student-friendly language | Clarify where learners are going | > Instruction: Rewrite the following formal learning objective in a more student-friendly way, something that is easier for students to understand and relate to.<br><br>Learning Objective: "…" | Simplified targets help students internalise expectations and self-monitor progress, a prerequisite for effective formative assessment. |
| Generate discussion questions for co-constructing success criteria | Clarify where learners are going | > I want to teach my class of undergraduate students in my Introduction to Functional Magnetic Resonance Imaging class to write a method section for an fMRI paper. They have already learned about the basics of fMRI experimental design and data analysis. Instead of reading out learning objectives, I want to co-construct learning objectives with students in my class. How do I do this? | Co-construction promotes ownership and transparency; LLM-seeded questions save planning time while fostering a classroom dialogue around goals. |
| Produce exemplar work at various quality levels | Clarify where learners are going | > I want to begin a session by having the class critique and improve sample methods sections. For this method section below describing experimental task, can you modify it to make it a very low quality one?<br><br>Method Section: "…" | Comparing and critiquing exemplars makes assessment criteria visible and refines students' concept of quality. |
| Automatic generation of diagnostic MCQs (or short-answer probes) | Assess where learners are in their learning | > Generate five multiple-choice questions for Grade 12 biology students to assess their foundational knowledge of DNA. The questions should focus on the structure and function of DNA, the role of nucleotides, and the difference between DNA and proteins. The questions should be factual. Include one correct answer and three plausible distractors for each question. These questions should not test and require knowledge of transcription, translation, and processes related to protein synthesis. Provide rationales for each distractor. | Rapid, curriculum-aligned questions reveal misconceptions before, during, or after instruction, letting teachers adapt teaching in short formative assessment cycles. |
| Task-level feedback on student work | Move learning forward | > Evaluate the essay paragraph below only for biological accuracy and completeness; highlight any factual errors and missing concepts in two bullet-points. | Immediate, specific task feedback corrects misunderstandings and closes knowledge gaps. |
| Process-level feedback (e.g., strategy hints) | Move learning forward | > Review the attached Python code. Suggest two alternative strategies the student could try to make the algorithm more efficient, without writing full code for them. | Guiding strategy, not just answers, cultivates transferable problem-solving skills and deeper learning. |
| Self-regulation feedback | Move learning forward | > Example zero-shot prompt template with structured output constraints : You are a metacognitive coach helping a student become a more self-regulated learner. Context: - Assignment type: {essay / lab report / code file / …} | Metacognitive prompts nurture students' ability to plan, monitor and evaluate their own learning. |

- Learning goals or rubric (concise bullets): { … }

- Student submission: <<< {paste text or code} >>>

Task:

1. DO NOT correct specific content or provide model answers.

2. Identify ONE strength in the student's approach.

3. Identify ONE area where the student could improve their strategy or mindset.

4. Give the student THREE reflective prompts or sentence stems that will help them:

   a. Evaluate their current approach

   b. Plan a concrete next step

   c. Monitor their progress while revising

Output in this format:

**Strength** → …

**Strategic growth area** → …

**Reflective prompts**

1. …

2. …

3. …

## 4.2 Developing Metrics to Evaluate LLM-Generated Feedback Across Different Feedback Types

While LLMs have been widely explored for generating feedback, much of the literature does not distinguish between different types of feedback and their unique roles in learning. Formative assessment relies on various feedback types—such as task-level feedback, process-level feedback, and self-regulation feedback—each serving distinct functions in guiding student progress (Hattie & Timperley, 2007). Educators should be intentional in determining which types of feedback LLMs can effectively enhance relevant academic outcomes.

On the other hand, evaluating the efficacy of LLM-generated feedback requires a structured approach that considers the different types of feedback essential to formative assessment. Existing research on LLM feedback primarily focuses on task-level feedback, which assesses the correctness of a specific task (Balse et al., 2023; Koutcheme et al., 2024; Leinonen et al., 2023; Sarsa et al., 2022). However, even within this commonly studied area, LLM applications remain limited to identifying errors and knowledge gaps rather than helping students understand why they may have misunderstood a concept. Furthermore, there is a lack of benchmarks that assess LLM's efficacy in generating different types of feedback, making it difficult to develop evidence-based LLM models grounded in pedagogical principles that can effectively enhance academic outcomes through targeted feedback. Without a structured evaluation framework, LLM-generated feedback risks being generic or misaligned with students' learning needs. To address this, it is essential to map out the nuances of task-level, process-level, and self-regulation feedback, while ensuring that LLM applications are developed with a clear focus on generating specific types of feedback aligned with their distinct instructional purposes. Furthermore, when designing such benchmarks, it is crucial to consider that all forms of formative feedback, including task-level feedback, should guide the "next steps" in instruction and learning. Clear and well-defined benchmarks will enable researchers and educators to assess how LLMs contribute to formative feedback and assessment in meaningful ways, ensuring that LLM-driven feedback moves beyond superficial evaluations to actively support student learning and conceptual understanding.

## 4.3 Using LLMs to Overcome Systemic and Cultural Barriers in Formative Assessments

Despite widespread recognition of the benefits of formative assessments, systemic barriers continue to hinder their effective implementation (Johan Krumsvik & Ludvigsen, 2013; Yorke, 2003), particularly in non-Western educational contexts (Al-Wassia et al., 2015;

Mok, 2011; Prompiengchai et al., 2024; J. Yang & Tan, 2019). Formative assessments are inherently student-centered, requiring active student engagement, self-reflection, and iterative feedback (Connell et al., 2016; Evans et al., 2014; Topping, 2005, 2009; Wood, 2009). However, many education systems, especially in countries with deeply ingrained teacher-centered models, struggle to adopt formative assessments at scale (Mok, 2011; Prompiengchai et al., 2024; Tay & Ng, 2015). In teacher-centered classrooms, instruction is predominantly lecture-based, with teachers serving as the primary source of knowledge while students passively receive information. Shifting toward student-centered learning may require cultural and structural changes, which can be difficult to implement despite policy efforts (Brinkmann, 2015, 2019). For example, while the Indian government acknowledges the importance of formative assessments in its National Education Policy (Ministry of Education, 1986; Ministry of Human Resource Development, 2020; National Council of Educational Research and Training, 2005), current ways of teaching and learning still rely heavily on rote memorization and high-stakes exams (Bhattacharya, 2022; Boucher, 2024), limiting the opportunities for dynamic, feedback-driven learning that is central to the successful implementation of formative assessments.

Several systemic barriers contribute to these challenges. Teachers often face overwhelming workloads, exacerbated by low pay and limited institutional support, reducing their motivation and capacity to adopt new teaching methods. Additionally, rigid examination systems, such as national entrance exams in countries like China and South Korea, prioritize high-stakes testing over continuous, formative feedback, leaving little room for iterative learning (Kirkpatrick & Zang, 2011; Tan & Yang, 2021; J. Yang & Tan, 2019). Moreover, many educators lack pedagogical training in formative assessments, making it difficult to design and implement effective strategies. On the student side, unfamiliarity with formative assessments can lead to resistance, as learners may not be accustomed to self-assessment, peer feedback, or open-ended assignments that encourage reflection and conceptual understanding (Mok, 2011; Prompiengchai et al., 2024). These barriers, both cultural and systemic, make it challenging to integrate formative assessments into traditional educational models.

However, large language models present a unique opportunity to address these challenges. One of the primary advantages of LLMs is their ability to save teachers time by automating aspects of formative assessments. As discussed in previous sections of this review, LLMs can generate personalized feedback, simplify learning objectives, and provide structured prompts that align with student-centered learning—all at no cost when using models like ChatGPT's free version. This makes it more feasible for teachers to integrate formative assessments into their instruction without significantly increasing their workload. For instance, teachers who struggle with designing student-friendly rubrics or guiding reflective discussions can use LLMs to scaffold these elements (Sridhar et al., 2023), ensuring alignment with formative assessment principles.

Additionally, LLMs could support teachers in developing hybrid lesson plans that integrate both traditional "chalk-and-talk" methods and student-centered formative assessment strategies. This idea of blending traditional instruction with student-centered approaches was suggested by an Indian teacher participant who implemented a peer-assessment technology for the first time, noting that a gradual transition—rather than an abrupt shift—could help educators and students adapt more effectively (Prompiengchai et al., 2024). For example, an AI-powered lesson-planning tool could help teachers embed formative assessments into their lectures by generating reflection prompts, peer discussion questions, and scaffolded exercises that gradually shift students toward active engagement.

Beyond classroom instruction, LLMs could also serve as personalized learning agents, providing students with low-cost access to formative feedback outside of class (Botarleanu et al., 2019). In regions where teacher availability is limited, LLMs could act as supplementary tutors, helping students engage with learning materials at their own pace while receiving automated guidance on their progress (Cao, 2023; Finnie-Ansley et al., 2022; Matelsky et al., 2023). By incorporating elements of formative assessments into these interactions—such as prompting students to explain their reasoning, guiding them through self-assessment exercises, and encouraging metacognitive reflection—LLMs can help foster a culture of continuous learning in diverse educational contexts. While LLMs are not a substitute for structural reforms in education (Karataş et al., 2025), they provide a practical and scalable tool to mitigate some of the key barriers preventing the widespread adoption of formative assessments. Research informing the development of LLM-driven technology should involve diverse stakeholders, including students and teachers, as equal partners throughout the process (Cleverley et al., 2025; Matthews et al., 2018; Ritter et al., 2024). Ultimately, LLMs have the potential to bridge systemic gaps in formative assessment, making student-centered learning more feasible across diverse educational contexts.

## 4.4 Rethinking Formative Assessments in an LLM-Powered Classroom

As large language models become more integrated into education, concerns have emerged about students relying on these tools to complete assessment tasks rather than engaging in meaningful learning (Kovari, 2025; Ray, 2023). Traditional assessment tasks that focus on factual recall or basic comprehension can be easily completed by AI, potentially diminishing their effectiveness in fostering meaningful learning. This highlights the need for educators to rethink formative assessment design by emphasizing creativity, critical thinking, and problem-solving—skills that AI cannot fully replicate.

To design formative assessments that are less susceptible to LLM-generated responses, educators can incorporate AI-awareness by encouraging students to critically engage with AI-generated content (Tu et al., 2023). For example, students can analyze LLM-generated responses, identify biases or errors, and propose improvements. A potential assessment task could involve evaluating AI-generated sample work both before and after a time period. As students gain a deeper understanding of the subject matter, they will be better equipped to critique and refine AI-generated content, demonstrating their mastery of key learning objectives.

Beyond AI-aware strategies, formative assessments can also be redesigned to emphasize interactive and multimodal engagement. One approach is to create tasks that require iterative dialogue, where students engage in real-time discussions with conversational agents or peers, defending and refining their arguments based on evolving input (J. Liu et al., 2024). Another strategy is to incorporate multimodal assignments, such as video presentations, collaborative whiteboard exercises, or creative design projects, which demand synthesis across different formats and media (Harun & Singh, 2024; Lim et al., 2022; Ryan et al., 2010). Additionally, educators can prompt LLMs to generate unique formative assessment ideas tailored to fostering higher-order thinking, such as open-ended case studies requiring students to apply concepts in unfamiliar contexts, or reflection-based tasks that ask students to critique and refine AI-generated responses.

While these strategies help ensure formative assessments remain meaningful in an AI-driven world, further exploration is needed to develop assessment formats that harness AI to deepen student engagement and understanding. Although LLMs can be viewed as a challenge to assessment integrity, educators can design tasks that leverage AI to promote critical reflection, problem-solving, and adaptability (Tu et al., 2023). By incorporating AI-aware strategies and interactive, cognitively demanding tasks, formative assessments can encourage students to develop skills that are not only essential for academic success but also transferable across diverse real-world contexts. Thoughtfully designed assessments will help students engage more deeply with the learning process, ensuring that AI serves as a catalyst for intellectual growth rather than a tool for superficial completion.

## 4.5 Future Directions

Thus far, we have examined recent advances in the application of large language models for formative assessment, highlighting both their potential and the limitations of existing studies. Moving forward, several important avenues for future research emerge that could enhance our understanding and effective use of LLMs in formative assessment contexts:

- **Lesson Planning and Objectives:**
  - How can LLM-generated learning objectives be effectively aligned with formative assessment principles beyond grammatical correctness or Bloom's taxonomy?
  - What strategies enable educators to adapt generic or template-based LLM-generated lesson plans to diverse educational and cultural contexts?
  - How might interactive, LLM-supported simulations help educators practice co-constructing success criteria and facilitate classroom implementation?
- **Sample Work Generation:**
  - What methodologies enable generic or domain-specialized LLMs to produce authentic, diverse, and discipline-specific sample work suitable for formative assessment practices?
  - How can educators leverage limitations or gaps in LLM-generated samples to enhance students' critical reflection and deepen their understanding of assessment criteria?
- **Automatic Question and Feedback Generation:**
  - What approaches improve the ability of generic and fine-tuned LLMs to consistently generate formative assessment multiple-choice questions (MCQs) with high-quality, contextually appropriate distractors?
  - How can research systematically evaluate LLM-generated feedback using robust, pedagogically grounded metrics that distinguish between task-level, process-level, and self-regulation-level feedback?
  - To what extent do LLM-generated feedback strategies enhance learners' self-regulated learning skills, motivation, and self-efficacy?
  - What specific linguistic features and prompt-engineering methods most effectively foster self-regulated learning through LLM-generated feedback?
- **Cultural and Systemic Barriers:**
  - How can LLM-enabled formative assessment tools be adapted to align with national curricula, examination pressures, and teacher-centered beliefs without diluting core formative assessment principles?

- How can LLMs effectively support teachers in blending traditional instructional methods (e.g., chalk-and-talk) with student-centered formative assessment practices?
- How can we ensure high-quality, LLM-driven formative feedback in non-English languages?
- How do different deployment models of LLMs (e.g., chat-based assistants, learning-management system, plug-ins, mobile apps) affect teacher workload in large-class contexts?
- How might AI-generated self-regulation prompts be tailored to varying cultural conceptions of autonomy, authority, and collaborative learning?
- What models of participatory research most effectively integrate diverse stakeholders (teachers, students, educational researchers) in the development, deployment, and evaluation of LLM-driven formative assessment tools and frameworks?

- **Standards and EdTech Procurement:**
  - How can LLM-driven formative-assessment tools be potentially validated against national or state standards frameworks?
  - What criteria should EdTech procurement teams in schools and universities include to compare LLM-enabled FA platforms (e.g., quality of feedback, evidence of pedagogical efficacy)?
  - How can open standards be extended or adapted to ensure that LLM-based formative assessment tools integrate smoothly with existing learning-management systems, analytics dashboards, and future ed-tech upgrades?

- **AI-Aware Formative Assessment Tasks:**
  - How can formative assessment tasks be re-designed to leverage LLM capabilities while encouraging deeper student engagement, critical thinking, and transferable skills?

## 5. Conclusion

In this review, we have systematically situated the emerging applications of large language models within the established theoretical framework of formative assessment and feedback (Black & Wiliam, 2009; Hattie & Timperley, 2007). By organizing existing LLM research around core principles of formative assessment—clarifying intended learning goals, accurately assessing student's knowledge, and facilitating meaningful feedback to inform instructional process and advance student learning—we clarified how researchers and educators can use LLMs to support and enhance formative assessment processes that result in fostering meaningful pedagogical outcomes. We utilized the GPS-guided learning journey as a metaphor to demonstrate how LLM can automate different components of formative assessment such as generating student-friendly learning objectives and delivering adaptive assessment and personalized feedback while emphasizing that LLMs can only work in conjunction with teacher's and student's judgment who decide how their learning journey should look like and when should AI be used (Figure 1). Through example prompts and outputs, we provide guiding principles for educators to effectively leverage cost-free LLMs in formative assessments while remaining grounded in pedagogical best practices. The hope is that educators are equipped with a basic understanding of formative assessments, LLMs, and prompting techniques, such that they gain the intuition to experiment with using generic LLMs such as ChatGPT to scale the pedagogically-informed implementation of formative assessments within their classrooms.

Furthermore, we extensively reviewed the use of large-language models in generating formative feedback and showed that much of the LLM studies evaluate feedback as a unitary component that is assumed to improve student learning. We pointed out limited research in specifically generating self-regulated feedback; though, LLM-based feedback in general may improve learners' self-regulated learning. Future research should focus explicitly on developing rigorous evaluative metrics that differentiate among various types of formative feedback, enabling educators and researchers to systematically assess and refine LLM-driven interventions.

Additionally, cost-free LLM has the potential to overcome current systemic and cultural barriers that hinder the effective implementation of formative assessment by saving teachers' time in designing formative assessments and scaling personalized feedback. We suggest that LLMs can reduce teachers' workload while also fostering additional pedagogical outcomes by incorporating formative assessment strategies into their existing teaching plans. It may be possible that LLMs can design lesson plans that are culturally relevant and reconcile with teacher's and student's philosophies of learning that would normally be at conflict with the constructivist pedagogy and student-centered principles of formative assessment. Ultimately, careful integration of LLMs into formative assessment practices offers a powerful opportunity to enhance educational practices, foster deeper student engagement, and cultivate critical, transferable skills—paving the way for broader implementation of formative assessment across diverse global educational settings.

# Supplementary Information

**A Practical Guide for Supporting Formative Assessment and Feedback Using Generative AI**

Sapolnach Prompiengchai[1,2,3], Charith Narreddy[3,4], Steve Joordens[2,3]

1. Department of Psychiatry, University of Oxford, Oxford, UK
2. Department of Psychology, University of Toronto Scarborough, Toronto, Canada
3. Clematis Research Empowerment Hub, Toronto, Canada
4. Denmark High School, Georgia, United States

This supplementary information contains the following:

- Appendix A: ChatGPT Response for Co-Constructing Learning Objectives with Students
- Appendix B: ChatGPT Response for Adding Unnecessary Details to fMRI Methods Section

**Appendix A: ChatGPT Response for Co-Constructing Learning Objectives with Students**

I want to teach my class of undergraduate students in my Introduction to Functional Magnetic Resonance Imaging class to write a method section for an fMRI paper. They have already learned about the basics of fMRI experimental design and data analysis. Instead of reading out learning objectives, I want to co-construct learning objectives with students in my class. How do I do this?

**ChatGPT's Response:**

Co-constructing learning objectives with your students is a great way to foster engagement and ensure that the objectives are relevant and meaningful to them. Here's a step-by-step process you can follow to co-construct learning objectives for your fMRI paper writing activity:

### 1. Introduce the Topic

Start by introducing the task at hand: writing a method section for an fMRI paper. You could briefly explain that this is a key skill for communicating research findings clearly and that it requires both a good understanding of the experimental design and the data analysis techniques they've already learned.

### 2. Prompt a Discussion

Ask open-ended questions to stimulate thinking and engage the class. The goal is to guide students to reflect on what they should know and be able to do when writing the methods section. Here are some questions to get you started:

- **What is the purpose of the methods section in a research paper?**
  - *This will help students realize the importance of clear and precise description of methods so others can replicate the study.*
- **What components should a methods section include when writing about fMRI experiments?**
  - *This will help them focus on critical elements like experimental design, participants, stimuli, fMRI acquisition, data analysis procedures, etc.*

- **What knowledge and skills do you think are necessary for writing a method section in an fMRI paper?**
  - *This will help them reflect on what they need to know about fMRI methodology, including how to describe scanning protocols, preprocessing, statistical analysis, and so on.*

## 3. Identify Key Themes

As students discuss the questions, jot down the key ideas on the board or a flip chart. After a few minutes of brainstorming, start organizing these ideas into categories or themes. For example:

- **Participants** (Who were the participants? How were they selected?)
- **fMRI Acquisition** (What scanner was used? What was the scanning protocol?)
- **Experimental Design** (What were the conditions, stimuli, or tasks? What was the counterbalancing?)
- **Data Analysis** (What preprocessing steps were taken? What statistical methods were used?)

## 4. Translate Ideas into Objectives

With the themes laid out, guide the class in turning these ideas into clear, actionable learning objectives. A good way to do this is by using the "I can" statement format, which will help students focus on specific skills. For example:

- **"I can describe the fMRI experimental design clearly and precisely."**
- **"I can explain the process of data acquisition, including the type of scanner and protocol used."**
- **"I can outline the preprocessing steps typically involved in fMRI data analysis."**
- **"I can provide a clear description of the statistical methods used in an fMRI study."**

Ask students if these objectives reflect what they feel is important, and allow them to make revisions or additions. This ensures the objectives are in line with their understanding and what they want to achieve.

## 5. Refine and Finalize

Once you have a list of objectives, review them with the class to ensure clarity and completeness. Encourage students to discuss if there's anything they think might be missing. If students suggest other skills or knowledge they need to accomplish the task, revise or add those to the list.

**6. Make the Objectives Concrete**

To ensure the objectives are measurable, ask students to clarify what it looks like when they've achieved them. For example:

- **"I can explain the preprocessing steps involved in fMRI data analysis"** might be followed up with an example, such as: *"I can list the preprocessing steps, including motion correction, normalization, and smoothing, and explain why each is important."*

**7. Review and Reinforce**

Once you have your finalized set of objectives, reinforce them throughout the class as you move into the more hands-on aspects of the lesson. When introducing tasks like writing a draft of a methods section, you can refer back to these objectives as a guide for what students should be aiming to achieve.

**Sample Learning Objectives for Writing a Method Section in an fMRI Paper**

By the end of this lesson, students should be able to:

- **Describe the experimental design of an fMRI study, including task types, stimuli, and counterbalancing methods.**
- **Provide a detailed description of the fMRI data acquisition process, including scanner specifications, sequences, and parameters.**
- **Outline the steps involved in preprocessing fMRI data, including motion correction, spatial normalization, and smoothing.**
- **Explain the statistical analysis methods used to analyze fMRI data, including statistical thresholds and models.**
- **Write a clear and concise methods section for an fMRI paper that would allow another researcher to replicate the study.**

**8. Wrap-Up**

Finally, end the discussion by summarizing the co-constructed objectives and reiterating how they will guide the next part of the class, where students will practice writing their method sections. You can tell students that by the end of the class, they should be able to write a complete method section that follows the objectives they co-constructed.

By co-constructing learning objectives in this way, you empower students to take ownership of their learning, and you ensure that the objectives reflect their own needs and expectations for the class.

## Appendix B: ChatGPT Response for Adding Unnecessary Details to fMRI Methods Section

<u>User's Input</u>

I want to teach my class of undergraduate students in my Introduction to Functional Magnetic Resonance Imaging class to write a method section for an fMRI paper. They have already learned about the basics of fMRI experimental design and data analysis. I want to begin a session by having the class critique and improve sample methods sections. For this method describing fMRI data pre-processing, can you add unnecessary details about the researchers clicking the buttons in the FSL graphical user interface?

<u>ChatGPT's Response</u>

Certainly! Here's a satirical version of an fMRI methods section, where unnecessary details about clicking buttons in FSL's graphical user interface are added. The point is to highlight how these kinds of overly specific descriptions can detract from the clarity and usefulness of a scientific methods section.

**Sample Method Section (with unnecessary details):**

**fMRI Data Preprocessing**

The raw fMRI data were first loaded into FSL (FMRIB Software Library) for preprocessing. The researchers, with the experiment data organized neatly on their desktop, opened FSL's main graphical user interface (GUI) by double-clicking on the FSL icon. A quick glance confirmed that all required tools were visible in the GUI's toolbar. They then navigated to the "FEAT" option, located under the "Analysis" tab, where they clicked once to open the analysis setup window. The mouse was carefully moved across the screen to select the appropriate folder containing the subject data.

Within the FEAT setup window, the researchers clicked the "Pre-statistics" tab. With the mouse pointer hovering over the tab for a brief moment, they selected the "Slice Timing Correction" option, ensuring the correct slice acquisition order was specified from the drop-down menu. It was noted that the "Interleaved" option was selected, as the participants' scan used this sequence. The "Motion Correction" button was then clicked with a deliberate movement of the cursor, followed by confirming the reference image was selected (the first image in the functional run), indicated by the checkmark in the GUI.

The researchers then proceeded to the "Spatial Filtering" section. Here, the mouse was carefully maneuvered to adjust the "Gaussian Kernel" parameter, selecting a 5mm kernel from the drop-down menu. The button labeled "Apply" was clicked, and the progress bar was carefully monitored to ensure that the kernel was applied properly to the data. After several seconds, the window displayed "Processing Complete," and a soft click was made on the "OK" button to confirm.

To further refine the preprocessing, the "High-pass filtering" section was accessed. The researcher clicked on the input box for the high-pass cutoff, typing "100" into the text field, and clicking the "Enter" key. The researcher then double-checked that the appropriate temporal filtering was applied, confirming this by glancing back at the "High-pass filter" section in the GUI.

Finally, once all preprocessing steps had been configured, the researcher clicked the "Go" button at the bottom of the FEAT window to begin the preprocessing pipeline. They watched as the program executed the steps one-by-one, occasionally glancing at the terminal window to confirm that no error messages appeared. At times, the mouse was moved to interact with the terminal window if additional user input was required (e.g., for certain motion correction options).

After preprocessing was completed, the resulting images were saved in the designated output folder. A quick confirmation was made that the output files had the expected names, including the subject identifier and the suffix for each preprocessed data type (e.g., "smoothed_func.nii.gz" for the smoothed functional images).

---

**Critique Points:**

- **Unnecessary Detail**: The description of clicking specific buttons in the graphical user interface (GUI), such as navigating between tabs and clicking on checkboxes or dropdown menus, is excessive. The focus should be on what steps were performed, not the exact sequence of button presses or the interface's specific elements.
- **Lack of Generalization**: The description should be generalized to avoid implying that every researcher follows the exact same set of steps or sequence. It's sufficient to state that preprocessing steps (e.g., slice timing correction, motion correction, spatial smoothing) were applied without detailing how each button was clicked.

- **Software-specific Details**: While mentioning software like FSL is necessary, the methods should not become overly dependent on describing the GUI navigation. Mentioning the use of FSL and the specific tools (e.g., FEAT) is enough, and details about interacting with the software interface could be omitted.

---

This method section, with excessive detail on button-clicking, is useful to show students how too much focus on the mechanics of the process can distract from the key methodological points. It's also a good example of how to streamline a methods section for clarity and brevity!